\documentclass[%
 reprint,
superscriptaddress,
 amsmath,amssymb,
 aps,
pre,
]{revtex4-2}

\usepackage{graphicx}
\usepackage{dcolumn}
\usepackage{bm}
\usepackage{multirow}
\usepackage{xcolor}
\usepackage{hyperref}
\usepackage{gensymb}

\begin{document}

\title{Eigenvalue-based micromagnetic analysis of switching in spin-torque-driven structures}

\author{Z. Lin}
\affiliation{Department of Electrical and Computer Engineering, University of California San Diego, La Jolla, California 92093, USA}
\affiliation{Center for Memory and Recording Research, La Jolla, California 92093, USA}
\affiliation{Materials Science and Engineering Program, University of California San Diego, La Jolla, California 92093, USA}
\author{I. Volvach}
\affiliation{Department of Electrical and Computer Engineering, University of California San Diego, La Jolla, California 92093, USA}
\affiliation{Center for Memory and Recording Research, La Jolla, California 92093, USA}
\affiliation{Materials Science and Engineering Program, University of California San Diego, La Jolla, California 92093, USA}
\author{X. Wang}
\affiliation{Department of Electrical and Computer Engineering, University of California San Diego, La Jolla, California 92093, USA}
\affiliation{Center for Memory and Recording Research, La Jolla, California 92093, USA}
\author{V. Lomakin}
 \email{vlomakin@eng.ucsd.edu}
\affiliation{Department of Electrical and Computer Engineering, University of California San Diego, La Jolla, California 92093, USA}
\affiliation{Center for Memory and Recording Research, La Jolla, California 92093, USA}


\begin{abstract}
We present an eigenvalue-based approach for studying the magnetization dynamics in magnetic nanostructures driven by spintronic excitations, such as spin transfer torque and spin orbit torque. The approach represents the system dynamics in terms of normal oscillation modes (eigenstates) with corresponding complex eigenfrequencies. The dynamics is driven by a small number of active eigenstates and often considering just a single eigenstate is sufficient. We develop a perturbation theory that provides semi-analytical dynamic solutions by using eigenstates for the case in the absence of damping and spintronic excitations as a basis. The approach provides important insights into dynamics in such systems and allows solving several difficulties in their modeling, such as extracting the switching current in magnetic random access memories (MRAM) and understanding switching mechanisms. We show that the presented approach directly predicts the critical switching current, i.e., switching current for an infinite time. {The approach also provides solutions for the switching dynamics allowing obtaining the switching current for a finite switching time, provided that the system symmetry is broken, e.g., by tilting the polarizer, so that switching by a finite pulse is possible.}
\end{abstract}

\maketitle


\section{\label{sec:introduction}Introduction}
Spintronic structures, which utilize effects of spin polarization to drive the magnetization, are envisioned for multiple applications \cite{1,2,3,4}. Particular examples are spin transfer torque (STT) and spin orbit torque (SOT) magnetic random access memories (MRAM)  \cite{3,4,5,6,7,8,9,10} (Fig. 1). Such devices are based on the magnetization switching between two equilibrium magnetization states, which occurs when spin torque overcomes the system magnetic damping. Switching starts from one of the equilibrium states as small oscillations that increase in their magnitude to result in large oscillations, which are followed by the magnetization reversal to the other equilibrium state. Structures of a small size can be approximated by a single spin because the magnetization motion is mostly coherent \cite{5,11,12}. In the single spin approximation, switching properties, such as the switching current and time as well as the switching trajectory, can be obtained analytically \cite{13}. The switching current is typically related to the energy barrier and the switching efficiency, viz. the ratio between the switching current and energy barrier, is {independent of} the structure size. On the other hand, for larger structures, which are greater than the exchange or domain wall length, the magnetization dynamics is non-uniform \cite{5,11}. It makes studying switching properties of such structures more complicated. Switching can follow different trajectories depending on the strength and time dependence of the current. The switching current is not directly related to the energy barrier and the switching efficiency is not a constant with respect to the structure size [7,11]. There are no rigorous analytical models predicting the switching parameters and one needs to resort to numerical simulations, which are based on solving the Landau Lifshitz Gilbert equation (LLGE). Obtaining switching parameters numerically can be complicated and non-reliable. For example, obtaining the critical switching current, viz. the current required for switching at infinite time, is accomplished by calculating the switching current at multiple switching times and extrapolating to infinite time assuming a linear dependence of the switching current versus the inverse time \cite{5,11,14}. However, while such an extrapolation is accurate and efficient for small structures, for which a single-spin approximation is valid, it may be inaccurate and slow for larger structures. Additionally, relying purely on brute-force numerical simulations does not provide a clear physical picture as to why a certain type of switching occurs and how it is related to the operational parameters.
\begin{figure}
\includegraphics[width=\linewidth]{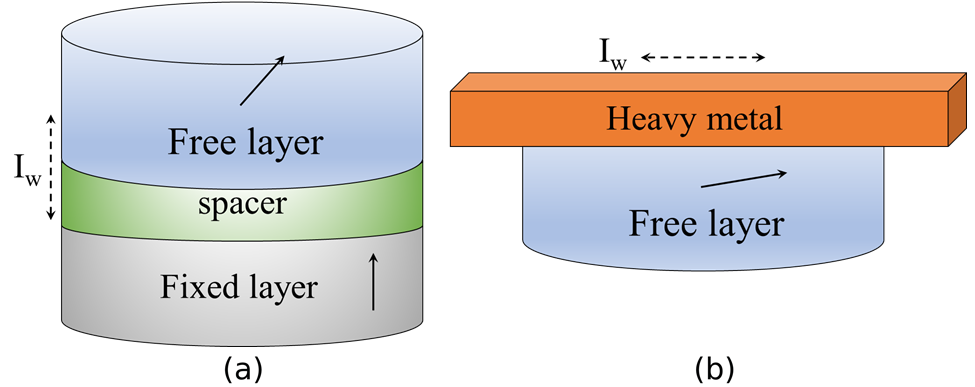}
\caption{Examples of spintronic structures. (a) STT-MRAM and (b) SOT-MRAM. $I_w$ denotes the flow of writing current.}
\end{figure}

Here, we present an eigenvalue framework (EVF) allowing for the study of switching properties in spintronic devices. The framework is based on using the linearized LLGE with effective fields including spintronic terms to find its eigenfrequency and corresponding eigenstate solutions. These solutions are used to represent the LLGE dynamic solutions as a sum over the damped or growing eigenstates modulated by time dependent coefficients for which simple equations are constructed. A perturbation theory is invoked to find the eigenfrequencies and eigenstates in the presence of spintronic excitations, allowing easily using the presented framework for cases with time dependent currents. The framework allows reliably obtaining the critical current density required for switching in spintronic devices as well as the switching current and switching time for pulsed current excitations. The framework also provides understanding of the switching dynamics behavior for structures of small and large sizes, including cases with non-uniform dynamics.

\section{\label{sec:methodology}Methodology}
The magnetization dynamics is described by the LLGE
\begin{equation}
\label{1}
\begin{split}
\frac{\partial \mathbf{m}}{\partial t}= &-{\gamma }'\mathbf{m}\times ({{\mathbf{H}}_{eff}}(\mathbf{m})+{{\mathbf{H}}_{stt}}(\mathbf{m})) \\
&-\alpha {\gamma }'\mathbf{m}\times \mathbf{m}\times ({{\mathbf{H}}_{eff}}(\mathbf{m})+{{\mathbf{H}}_{stt}}(\mathbf{m})) 
\end{split},
\end{equation}
where $\mathbf{m}$ is the normalized magnetization, ${\gamma }'={\gamma }/{(1+{{\alpha }^{2}})}$ is the modified gyromagnetic ratio, $\gamma$ is the gyromagnetic ratio, and $\alpha$ is is the damping constant. The term ${{\mathbf{H}}_{eff}}(\mathbf{m})$   is the effective field, which includes the magnetostatic field ${{\mathbf{H}}_{ms}}$, exchange field ${{\mathbf{H}}_{ex}}$, and and anisotropy field (assumed uniaxial) ${{\mathbf{H}}_{an}}$:
\begin{equation}
\label{2}
\begin{split}
  & {{\mathbf{H}}_{eff}}={{\mathbf{H}}_{ms}}+{{\mathbf{H}}_{ex}}+{{\mathbf{H}}_{an}}=C\mathbf{m}, \\ 
 & {{\mathbf{H}}_{ms}}={{M}_{s}}\nabla \int{\frac{\nabla \cdot \mathbf{m}}{|\mathbf{r}-\mathbf{{r}'}|}}d\mathbf{{r}'};\,\,\,{{\mathbf{H}}_{ex}}=\frac{2A}{{{M}_{s}}}{{\nabla }^{2}}\mathbf{m};\\
 &{{\mathbf{H}}_{an}}={{\frac{2K_U}{M_s}}}(\mathbf{\hat{k}}\cdot \mathbf{m})\mathbf{\hat{k}}, \\ 
\end{split}
\end{equation}
where $M_s$ is the saturation magnetization, $A$ is the exchange constant, {$K_U$ is the anisotropy energy density,} $\mathbf{\hat{k}}$ is the uniaxial anisotropy axis direction. The effective field is linear in $\mathbf{m}$ and, therefore, $C$ is the linear field operator that is {independent of} $\mathbf{m}$. In Eq. (1), the term ${{\mathbf{H}}_{stt}}(\mathbf{m})$ is the spin transfer torque field that using simplified models can be given by
\begin{equation}
\label{3}
{{\mathbf{H}}_{stt}}=\beta \mathbf{m}\times \mathbf{p}.
\end{equation}
Here, $\mathbf{p}$ is the polarization direction and $\beta$ is an STT parameter that can be written as $\beta=Jb$, where $J$ is the electric current density and $b$ is a  coefficient related to the spintronic excitation. For magnetic tunnel junctions (MTJs) or spin valves, the coefficient $b={\eta {{\hbar }}}/{(2e{{M}_{s}}\delta )}$, where $\eta$ is the spin transfer torque efficiency, $\hbar$ is reduced Plank’s constant, $e$ is the electron charge, and $\delta$ is the effective thickness \cite{15,16,17}. For the spin Hall effect, $b={{{\mu }_{B}}{{\theta }_{SH}}}/{e{{M}_{s}}\delta }$ where ${\theta }_{SH}$ is the spin Hall angle, and ${\mu }_{B}$ is the Bohr magnetron \cite{18,19}. We note that $\beta$ and $\mathbf{p}$ can be functions of space and time, e.g., the current $J$ can be a pulse used for switching an MRAM cell. Additional effective field components can be added to ${{\mathbf{H}}_{eff}}(\mathbf{m})$ in Eq. (2), such as applied and magnetostrictive fields.

The LLGE (1) is non-linear in $\mathbf{m}$ due to the presence of the cross products and it describes the magnetization dynamics in a broad range of situations, including linear and non-linear effects. In many cases, however, the general LLGE can be linearized. Such a linearization is allowed when the magnetization varies only slightly from its equilibrium state. Weak magnetization variations can be due to weak excitations, e.g., by weak applied fields or by STT. The variations are also weak in the initial stages of the magnetization dynamics near an equilibrium state even when the system is driven by strong STT. The initial dynamics contains important information about the system behavior. In this section, we present a framework that uses a linearized LLGE to characterize and predict the switching behavior of magnetic devices under the effects of STT. We first present a linearized time domain LLGE. Then, we show the construction of eigen-solutions to study small oscillations around the equilibrium magnetization state in magnetic systems with an arbitrary geometry, damping factor, and STT. We, then, derive a method to solve time domain evolution based on the eigen-solutions.

\subsection{\label{sec:2.1}Linearized LLGE}
We define the equilibrium magnetization state ${{\mathbf{m}}_{0}}$  for the system without an STT field. This equilibrium is given by the Brown condition \cite{20}:
\begin{equation}
{{\mathbf{m}}_{0}}\times {{\mathbf{H}}_{eff}}({{\mathbf{m}}_{0}})=0,
\end{equation}
which corresponds to ${\partial {{\mathbf{m}}_{0}}}/{\partial t}=0$  when $\mathbf{H}_{stt}$ is excluded in Eq. (2). The effect of the system excitation by STT can be considered as a perturbation from this equilibrium.

We seek a solution for small magnetization deviations $\mathbf{v}$  around the equilibrium state such that
\begin{equation}
\mathbf{m}={{\mathbf{m}}_{0}}+\mathbf{v},
\end{equation}
and $\mathbf{v}$ is normal to $\mathbf{m}_{0}$, so that the normalization of $\mathbf{m}$ is maintained. Because of the linearity of ${{\mathbf{H}}_{eff}}(\mathbf{m})$, we can write ${{\mathbf{H}}_{eff}}(\mathbf{m})={{\mathbf{H}}_{0}}+C\mathbf{v}$, where ${{\mathbf{H}}_{0}}={{\mathbf{H}}_{eff}}({{\mathbf{m}}_{0}})$. Since $\mathbf{v}$ is normal to $\mathbf{m}_{0}$, we can project every operator into the tangent space $TM({{\mathbf{m}}_{\mathbf{0}}})$ by using the projection operator \cite{21}
\begin{equation}
{{P}_{{{\mathbf{m}}_{\mathbf{0}}}}}=(I-{{\mathbf{m}}_{\mathbf{0}}}\otimes {{\mathbf{m}}_{\mathbf{0}}}).
\end{equation}
Using Eq. (5) in Eq. (1), assuming that the damping constant $\alpha$ is small, which is the case for materials for which STT excitations are typically used, denoting the cross operator as $\Lambda (\mathbf{u})\mathbf{v}=\mathbf{u}\times \mathbf{v}$ and keeping only the terms linear in $\mathbf{v}$ and $\alpha$, we can write a linearized LLGE for $\mathbf{v}$:
\begin{equation}
\frac{\partial \,\mathbf{v}}{\partial t}=A\mathbf{v}-\gamma \beta \Lambda ({{\mathbf{m}}_{0}})\Lambda ({{\mathbf{m}}_{0}})\mathbf{p}.
\end{equation}
Here, the linear operator $A$ is defined as 
\begin{equation}
A=\Lambda ({{\mathbf{m}}_{0}}){{A}_{0\bot }}+\Lambda ({{\mathbf{m}}_{0}})\left( \alpha \Lambda ({{\mathbf{m}}_{0}}){{A}_{0\bot }}-\gamma \beta {{\Lambda }_{\bot }}(\mathbf{p}) \right),
\end{equation}
where
\begin{equation}
\begin{aligned}
&{{A}_{0\bot }}={{P}_{{{\mathbf{m}}_{\mathbf{0}}}}}{{A}_{0}}={{P}_{{{\mathbf{m}}_{\mathbf{0}}}}}\left( ({{\mathbf{H}}_{0}}\cdot {{\mathbf{m}}_{0}})I-C \right)\\
&\Lambda_{\bot}(\mathbf{p})={P}_{{{\mathbf{m}}_{\mathbf{0}}}}\Lambda(\mathbf{p})
\end{aligned},
\end{equation}
with unit operator $I$. One can prove that, when restricted to vector fields in $TM({{\mathbf{m}}_{\mathbf{0}}})$, the operator  $\Lambda ({{\mathbf{m}}_{0}})$ is linear and anti-symmetric and it is also invertible, i.e.,
\begin{equation}
\Lambda ({{\mathbf{m}}_{0}})\Lambda ({{\mathbf{m}}_{0}})=-I.
\end{equation}
In Eq. (7) combined with Eq. (8), the first term in the right hand side is linear in $\mathbf{v}$ and it corresponds to the precessional torque of the effective field. The second term is also linear in $\mathbf{v}$ and it corresponds to the damping torque of the effective field and the linear component of the STT torque. The last term of Eq. (7) can be regarded as a forcing term corresponding to the STT torque, which is a term independent of $\mathbf{v}$. The magnetization dynamics via the linearized LLGE can be solved numerically using standard finite difference or finite element methods \cite{22,23,24}, in which the solution is for the magnetization given as a set of points, e.g., centers of bricks in the finite difference methods or vertices of tetrahedrons in tetrahedral mesh based finite elements methods. Here, we show that the solutions can also be obtained by representing the magnetization as a superposition of eigenstates.

\subsection{\label{sec:2.2}Eigenvalue problem and perturbation analysis}
Based on the linearized LLGE (7), we can set up a linear eigenvalue problem:
\begin{equation}
A{{\varphi }_{n}}=j{{\omega }_{n}}{{\varphi }_{n}},
\end{equation}
for the complex valued small deviation magnetization eigen-states $\varphi_{n}$ and complex eigenfrequencies $\omega_{n}$. Here, $j=\sqrt{-1}$ is the imaginary unit. We recall that $\beta$ and $\mathbf{p}$ in Eqs. (3) and (7) can be functions of space and time and, therefore, the eigenvalue problem in Eq. (11) is defined at a particular time $t$.

The eigenstates and eigenfrequencies can be obtained analytically for some problems, e.g., in a single-spin approximation or numerically for general problems. The numerical solutions, e.g., can be based on finite difference and finite element methods similar to solutions of general micromagnetic problems \cite{22, 23, 24}. To present important properties of the eigen-solutions and provide a practical method for solving the time domain linearized LLGE (Eq. (7)), we present a perturbation solution of the eigenvalue problem of Eq. (11).

For the perturbation solution, we recall that we consider small damping cases with  $\alpha \ll 1$. We note that the parameter $\beta$ of the STT terms is usually on the same order as $\alpha$ \cite{17}, i.e., the linear STT term is small as well. Therefore, we can carry out a perturbation theory in which we define the base eigenvalue problem:
\begin{equation}
\Lambda ({{\mathbf{m}}_{0}}){{A}_{0\bot }}{{\bar{\varphi }}_{n}}=j{{\bar{\omega }}_{n}}{{\bar{\varphi }}_{n}},
\end{equation}
which has eigenfrequencies ${{\bar{\omega }}_{n}}$ and eigenstates ${{\bar{\varphi }}_{n}}$. The eigenfrequencies ${{\bar{\omega }}_{n}}$  can be shown to be purely real and the eigenstates  ${{\bar{\varphi }}_{n}}$, when normalized, can be shown to satisfy the weighted orthonormality condition \cite{21}:
\begin{equation}
<{{\bar{\varphi }}_{n}},{{A}_{0\bot }}{{\bar{\varphi }}_{{{n}'}}}>_{\Omega}=\frac{1}{{{V}_{\Omega }}}\int\limits_{\Omega}{\bar{\varphi }_{n}^{*}{{A}_{0\bot }}{{{\bar{\varphi }}}_{n'}}dV}={{\delta }_{n{n}'}},
\end{equation}
where $<\cdot ,\,\,\,\cdot >_{\Omega}$  is the inner product defined as the integral over the entire domain $\Omega$ of the magnetic structure, the asterisk denotes the complex conjugation, and ${{\delta }_{n{n}'}}$ is the Kronecker’s symbol. Unlike the original eigenvalue problem of Eq. (11), the base eigenvalue problem of Eq. (12) is time independent. The eigen-states ${{\bar{\varphi }}_{n}}$ form an orthonormal basis that can be used to represent more general eigen-solutions and time domain solutions.

We recognize that the operator $A$ in the eigenvalue problem (11) can be written as $A=\Lambda ({{\mathbf{m}}_{0}}){{A}_{0}}+\delta A$, where
\begin{equation}
\delta A=\Lambda ({{\mathbf{m}}_{0}})\left( \alpha \Lambda ({{\mathbf{m}}_{0}}){{A}_{0\bot }}-\gamma \beta {{\Lambda }_{\bot }}(\mathbf{p}) \right),
\end{equation}
is the perturbation operator that has a much smaller norm that the base operator $\Lambda ({{\mathbf{m}}_{0}}){{A}_{0}}$. The eigenfrequencies and eigenstates of the original eigenvalue problem are obtained as ${{\omega }_{n}}={{\bar{\omega }}_{n}}+\delta {{\omega }_{n}}$ and  ${{\varphi }_{n}}={{\bar{\varphi }}_{n}}+\delta {{\varphi }_{n}}$. Following the perturbation analysis, keeping only the linear terms in the perturbations of the operators and solutions, and using the anti-symmetric property of $\Lambda(\mathbf{m_0})$ in Eq. (10), the perturbation to the eigenfrequency is given by
\begin{equation}
{\delta {{{\omega }}_{n}}}={{\bar{\omega }}_{n}}<{{\bar{\varphi }}_{n}},\gamma \beta {{\Lambda }_{\bot }}(\mathbf{p}){{\bar{\varphi }}_{n}}>_{\Omega}+j\alpha {{\bar{\omega }}_{n}}^{2}<{{\bar{\varphi }}_{n}},{{\bar{\varphi }}_{n}}>_{\Omega}.
\end{equation}
{Note that $\delta\omega_{n}$ is purely imaginary, which is due to the fact that $\bar{\omega}_{n}$ and $<{{\bar{\varphi }}_{n}},{{\bar{\varphi }}_{n}}>$ are real, and the real and imaginary parts of the complex vectors $\bar{\varphi }$ and $\gamma \beta {{\Lambda }_{\bot }}(\mathbf{p}){{\bar{\varphi }}_{n}}$ are perpendicular to each other resulting in a purely imaginary $<{{\bar{\varphi }}_{n}},\gamma \beta {{\Lambda }_{\bot }}(\mathbf{p}){{\bar{\varphi }}_{n}}>$.} As a result, we understand that the eigen-frequencies are complex, i.e.,
\begin{equation}
{{\omega }_{n}}={\omega }^{'}_{n}+j{\omega }^{''}_{n},
\end{equation}
where ${\omega }^{'}_{n}=\operatorname{Re}\{{{\omega }_{n}}\}$ and ${\omega }^{''}_{n}=\operatorname{Im}\{{{\omega }_{n}}\}$. In the perturbation approximation ${\omega }^{'}_{n}={{\bar{\omega }}_{n}}$  and  {${\omega }^{''}_{n}=-j\delta {{\omega }_{n}}$}. When $J=0$, i.e., with  $\beta=0$, it can be shown that ${\omega }^{''}_{n}>0$  for $\alpha>0$. For  $J<0$, i.e. $\beta<0$, ${\omega }^{''}_{n}$ has an even greater positive value. For $J>0$, i.e., for $\beta>0$, the positive value of ${\omega }^{''}_{n}$ decreases and there is a certain critical value of $J_{cn}$ for which ${\omega }^{''}_{n}=0$. At values  $J>J_{cn}$, ${\omega }^{''}_{n}<0$, which corresponds to increasing precessional amplitude as discussed in connection with the time dynamics in Sec. \ref{sec:3}. Similarly, the perturbation to the eigenstates is given by
\begin{equation}
\begin{split}
  & \delta {{{\bar{\varphi }}}_{n}}=\sum\limits_{m}{{{\varepsilon }_{nm}}{{{\bar{\varphi }}}_{n}}} \\ 
 & {{\varepsilon }_{nm}}=\frac{{{{\bar{\omega }}}_{m}}<{{{\bar{\varphi }}}_{n}},(j\alpha {{{\bar{\omega }}}_{m}}I+\gamma \beta {{\Lambda }_{\bot }}(\mathbf{p})){{{\bar{\varphi }}}_{m}}\,>_{\Omega}}{{{{\bar{\omega }}}_{n}}-{{{\bar{\omega }}}_{m}}},m\ne n \\ 
 & {{\varepsilon }_{nn}}=\frac{1}{2}<{{{\bar{\varphi }}}_{n}},(\gamma \beta {{\Lambda }_{\bot }}(\mathbf{p})+j\alpha {{{\bar{\omega }}}_{n}}I){{{\bar{\varphi }}}_{n}}\,>_{\Omega} \\ 
\end{split}.
\end{equation}

\subsection{\label{sec:2.3}Time domain solutions}
The eigen-solutions in Sec. \ref{sec:2.2} can be used to represent the solutions of the time domain problem of Eq. (7). To that end, we write $\mathbf{v}$ as
\begin{equation}
\mathbf{v}\approx 2\sum\limits_{n}{\operatorname{Re}\left\{ {{a}_{n}}{{{\bar{\varphi }}}_{n}} \right\}},
\end{equation}
i.e., it given in terms of the base eigenstates $\bar{\varphi}_n$, complex eigenfrequencies  $\omega_n$, and coefficients $a_n$  determining the excitation of the eigenstates.  The factor of 2 accounts for the fact that the two eigenstates are symmetric in the positive and negative frequencies. The base eigenstates $\bar{\varphi}_{n}$ are used instead of the actual eigenstates $\varphi_n$ assuming that the differences between $\bar{\varphi}_{n}$  and  $\varphi_n$ are insignificant, which is the case under the assumption of small $\alpha$. Similar approximations were used in related applications of eigenvalue based solutions in micromagnetics \cite{21,25} and other areas of physics, e.g., electromagnetics \cite{26}. Using $\bar{\varphi}_{n}$ has important benefits due to the fact that $\bar{\varphi}_{n}$ are time independent and have the orthogonality property of Eq. (13).

Substituting the representation of Eq. (18) into Eq. (7), using the eigenvalue problem of  Eq. (11) with the perturbation solutions of Eqs. (12)-(15), and the orthogonality in Eq. (13), and performing a weighted inner product with ${{A}_{0}}{{\bar{\varphi }}_{n}}(\mathbf{r})$ in both sides of Eq. (18), we obtain the following set of independent time domain differential equations for $a_n$:
\begin{equation}
\frac{d{{a}_{n}}}{dt}=j{{\omega }_{n}}{{a}_{n}}+P_{n}^{stt},
\end{equation}
where
\begin{equation}
P_{n}^{stt}=<\gamma \beta \Lambda ({{\mathbf{m}}_{0}})\Lambda ({{\mathbf{m}}_{0}})\mathbf{p},{{A}_{0}}{{\bar{\varphi }}_{n}}>_{\Omega}.
\end{equation}
Solution for $a_n$  can be given by analytically solving the ordinary differential Eq. (19) as
\begin{equation}
{{a}_{n}}(t)={{e}^{\int_{0}^{t}{j{{\omega }_{n}}(\tau )d\tau }}}\left[ {{a}_{n}}(0)+\int_{0}^{t}{{{e}^{\int_{0}^{\tau }{-j{{\omega }_{n}}({t}')d{t}'}}}}P_{n}^{stt}(\tau )d\tau  \right].
\end{equation}
Here, $a_n(0)$ is determined from the initial condition as  ${{a}_{n}}(0)=<\mathbf{v}(t=0),{{A}_{0}}{{\bar{\varphi }}_{n}}>$, where $\mathbf{v}(t=0)$  is the initial complex small deviation magnetization state, and the integrals in the power exponentials appear because $\omega_n$  is generally a complex time dependent function. The solutions of Eq. (21) are valid for any time dependence of the current, including constant and pulsed currents.

We note that the  EVF can be extended to include finite temperature effects by adding a stochastic thermal term in the right hand side of  Eq. (19), which can be given following the formulation leading to $P^{stt}_{n}$ in Eq. (20). {This extended EVF can be used to study write error rates in STT MRAM caused by finite temperature effects. It would require calculating $a_n(t)$  for many realizations of thermal noise, which is much faster than a similar number of full micromagnetic simulations.}

\section{\label{sec:3}Solution analysis}
An important case that provides fundamental device parameters is the case of a constant current density. An important parameter is the critical current density $J_c$, which is the current $J$ required for switching the magnetization between two equilibrium states over an infinite time. In the constant current case,  $\omega_n$ is independent of time and Eq. (21) simplifies to
\begin{equation}
{{a}_{n}}(t)={{e}^{-{\omega }^{''}_{n}t}}{{e}^{j{\omega }^{'}_{n}t}}\left( {{a}_{n}}(0)+\frac{P_{n}^{stt}}{j{{\omega }_{n}}} \right)-\frac{P_{n}^{stt}}{j{{\omega }_{n}}},
\end{equation}
where we wrote $\omega_n$  explicitly in terms of its real and imaginary parts. Noting that  ${\omega }^{''}_{n}$ can be positive or negative and based on Eq. (15),  one concludes that the small deviation magnetization $\mathbf{v}$ is given as a sum over damped or growing oscillations. When ${\omega }^{''}_{n}>0$ for all $n$, all the terms in the sum are decaying and no switching occurs. When, however, ${\omega }^{''}_{n}<0$ for one or more of $n$, the corresponding time domain eigenstate contributions grow in time exponentially, eventually making $\mathbf{v}$ large enough, such that non-linear effects start taking place and switching is obtained. As we mentioned earlier, ${\omega }^{''}_{n}>0$ when $J<J_{cn}$ and ${\omega }^{''}_{n}$ decreased with an increase of $J$.

Using the perturbation analysis result of Eq. (15), and assuming a spatially constant $J$, we can find a condition for $J_{cn}$ by setting ${\omega }^{''}_{n}=0$:
\begin{equation}
{{J}_{cn}}=\frac{-j\alpha {{{\bar{\omega }}}_{n}}<{{{\bar{\varphi }}}_{n}},{{{\bar{\varphi }}}_{n}}>_{\Omega}}{\gamma b<{{{\bar{\varphi }}}_{n}},{{\Lambda }_{\bot }}(\mathbf{p}){{{\bar{\varphi }}}_{n}}>_{\Omega}}.
\end{equation}
Using Eqs. (15) with (13), (14), and (23), and under the assumption of constant $J$ and  $\mathbf{p}$, we can write an expression for ${\omega }^{''}_{n}$ in the following convenient form
\begin{equation}
{\omega }^{''}_{n}=-\alpha {\omega }^{'}_{n}\left( \frac{J}{{{J}_{cn}}}-1 \right),
\end{equation}
where, ${\omega }^{''}_{n}$ is given only in terms of ${\omega }^{'}_{n}$ and $J_{cn}$ for any given $J$ and $\alpha$. 

We can define the critical current density $J_c$ as ${{J}_{c}}={{\min }_{n}}\{{{J}_{cn}}\}$ obtained for the eigenstate number $n_c$. At $J=J_c$, ${{\omega }^{''}_{{{n}_{c}}}}=0$, so that the STT effects overcomes the effect of the system damping, and any $J>J_c$ lead to increased oscillations and switching. In terms of the time dependence, the critical current density $J_c$ is the current density that leads to switching over the period of an infinite time. The ability to obtain $J_c$ by solving a single eigenvalue problem followed by the perturbation analysis is important for understanding the physical behavior of the structure as well as for simulation and design purposes. An available alternative is solving the LLGE (1) to find the switching current for a set of finite times and extrapolating to an infinite time \cite{14}. Such an approach is slow and leads to uncertainties in the result as shown in Sec. \ref{sec:4}.

The presented theory provides not only the critical current density but also an approximation for the switching current density for a given time or stitching time for a given current density $J>J_c$. To that end, {we can set a condition on $|{{a}_{n}}|$ to be at a certain level $|a_n|_{\max}$ to lead to switching. This can be set by requiring that $|{{a}_{n}}{{|}_{\max}}|\bar{\varphi}_n|_{\max}=\zeta$, where $\zeta$ is a constant of $O(1)$ and {$|\bar{\varphi}_n|_{\max}$ is the largest magnitude of the eigenstate $\bar{\varphi}_n$ in the magnetic domain of interest.}} Then, using Eqs. (22)-(24), we can obtain the conditions for the switching current $J_{sw}$ for a given pulse duration $\tau$ and switching time $t_{sw}$ for a given current $J>J_c$:
\begin{equation}
  \frac{{{J}_{sw}}}{{{J}_{c}}}=1+\frac{\log \left( \xi  \right)}{\alpha {{\omega }^{'}_{c}}\tau };\;  {{t}_{sw}}=\frac{\log \left( \xi  \right)}{\alpha {{\omega }^{'}_{c}}\left( {J}/{{{J}_{c}}}\;-1 \right)},
\end{equation}
where {$\xi \approx {\zeta }/{(|{{{\bar{\varphi }}}_{nc}}{{|}_{\max }}|{{a}_{{{n}_{c}}}}(0)+{P_{{{n}_{c}}}^{stt}}/{j{{\omega }_{{{n}_{c}}}}}|)}$ is a coefficient related to the initial magnetization conditions and the driving term.} For zero-temperature simulations starting from equilibrium, ${{a}_{{{n}_{c}}}}(0)=0$ and $\xi$ is determined by $P_{{{n}_{c}}}^{stt}$. For finite-temperature simulations  $\xi$ is mostly determined by the initial condition ${{a}_{{{n}_{c}}}}(0)$, which is related to the magnetization distribution caused by the thermal fluctuations.

\section{\label{sec:4}Result and Discussion}
We implemented the EVF as a part of the finite element method based micromagnetic simulator FastMag  \cite{22}, which can run on multi-core CPUs and GPUs, and allows efficiently handling highly complex problems. The numerical eigenvalue problem is solved with an iterative implicitly restarted and preconditioned Arnoldi method \cite{27}.

We demonstrate the EVF by considering switching in an MRAM cell, comprised of a cylindrical free layer of 1 nm thickness and two diameters ($D$) of 20 nm and 80 nm with {${{M}_{s}}=960\,\text{emu/c}{{\text{m}}^{\text{3}}}$, ${{A}_{ex}}=1\,\mu \text{erg/cm}$, $\alpha=-0.01$ and perpendicular anisotropy of ${{K}_{U}}=6.11\text{ Merg/c}{{\text{m}}^{\text{3}}}$}. An STT field acts at the bottom surface of the free layer {and the polarization direction is tilted with a small angle of ${0.1\degree}$ with respect to the perpendicular direction, i.e., $\mathbf{p}=\left( 0,\sin ({\pi/1800}),\cos ({\pi/1800}) \right)$. The small tilt in $\mathbf{p}$ is set to break the symmetry to result in non-vanishing values of $P_{n}^{stt}$.}
\begin{figure*}[htb]
\includegraphics[width=\linewidth]{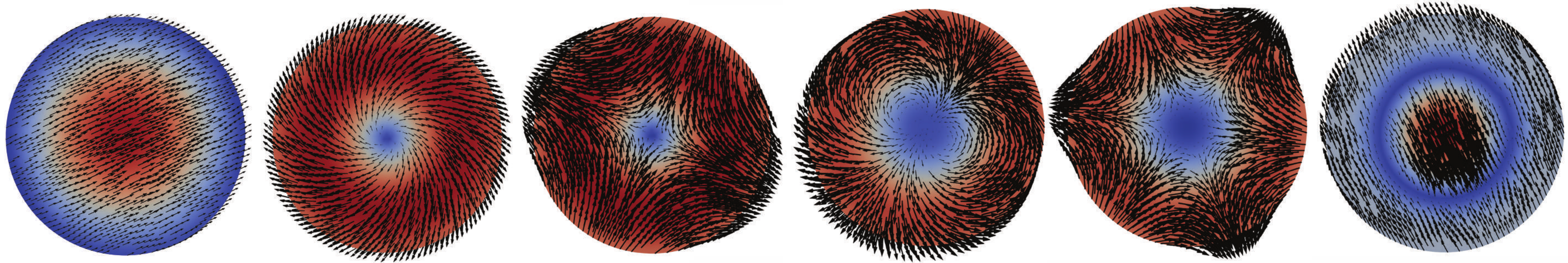}
\caption{First six eigenstates for 80 nm MRAM free layer disc. The color plot represents the magnitude of {eigenstates $|\bar{\varphi}_{n}|$}  and the arrow plot represents the real part of the eigenstate.}
\end{figure*}

 We start by solving the base problem as defined in Eq. (12). Fig. 2 shows the first six eigenstates of an 80 nm diameter free layer of a perpendicular MTJ. Table 1 provides the corresponding eigenfrequencies. The first eigenstate has a more uniform distribution with the maximum in the middle. The rest of the modes have a more non-uniform distribution with maxima and minima modulation. {The modes can be classified according to the winding numbers \cite{29}; here, the first six modes have winding numbers of 0, 1, -1, 2, -2, 1, respectively.} We also calculated the eigenstates of a smaller, 20 nm diameter, MRAM cell and found that its eigenstates have an almost the same spatial distribution. On the other hand, the eigenfrequencies of the 80 nm and 20 nm cells are different. The eigenfrequencies of the 20 nm cell are higher and have a much greater separation for different $n$.

Table 1 also shows $J_{cn}$, {and scaled $P^{stt}_{n}$ calculated with $J=J_{c1}$} for 20 nm and 80 nm cells. It is found that there is a nearly linear relation between $J_{cn}$ and ${\omega }^{'}_{n}$. To explain this behavior, we note that, using Eq. (23), $J_{cn}$ can be written as ${{J}_{cn}}={{k}_{n}}{\omega }^{'}_{n}$ , where $k_n$  is a coefficient related to the eigenstates. For the considered cases, $k_n$ is found to be nearly the same for all presented $n$. It follows that for a constant current, $J_{cn}$ is the smallest for the smallest ${\omega }^{'}_{n}$, i.e., the critical switching current $J_c$ corresponds to the smallest ${\omega }^{'}_{n}$. For small (20 nm) cells, the separation between  $J_{cn}$ is large, just like the separation between ${\omega }^{'}_{n}$, and, therefore, the contribution of the higher-order eigenstates is weak. For larger (80 nm) cells the separation between $J_{cn}$ is smaller and higher-order eigenstates may be excited. {For the small (20 nm) cells, the scaled $P_1^{stt}$ is greater than $P_n^{stt}$  for $n>1$, which again indicates that only the $n=1$ eigenstate can be strongly excited. For the larger (80 nm) cells, scaled $P_2^{stt}$ has a similar value as $P_1^{stt}$, and the other $P_n^{stt}$ have values greater than those for the 20 nm cells, which also indicates that multiple eigenstates can be excited.}
\begin{table}
\caption{\label{tab:table1} {EVF parameters: Eigenstate \# $n$, $f_n^{'}=\omega_n^{'}/(2\pi)$, $J_{cn}$, and $P^{stt}_{n}$ calculated with $J=J_{c1}$ and scaled with ${|{{\varphi }_{n}}{{|}_{\max }}}/{{{\omega }_{n}}}$ to make it unitless and related to the magnetization values.}}
\begin{ruledtabular}
\begin{tabular}{ccccc}
\textrm{$D$, [nm]}&
\textrm{$n$}&
\textrm{$f_{n}^{'}$, [GHz]}&
\textrm{$J_{cn}$, [MA/cm$^2$]} &
{\textrm{$|{{{P_{n}^{stt}{{\varphi }_{n}}}/{{{\omega }_{n}}|}\;}_{\max }}$}}\\
\colrule
\multirow{6}{*}{20}&1 &8.13 & 0.85 & {7.57e-7}\\
                   &2 & 29.71 & 3.10 & {3.94e-7}\\
                         &3 & 29.81	& 3.11 & {1.99e-7}\\
                         &4 & 65.55	& 6.82 & {7.61e-8}\\
                         &5 & 65.61 & 6.83 & {1.03e-7}\\
                         &6 & 94.18	& 9.93 & {2.78e-8}\\
\colrule
\multirow{6}{*}{80}&1 &3.76 &	0.39 & {1.01e-6}\\
                   &2 & 5.80 &	0.60 & {1.08e-6}\\
                         &3 & 5.86 &	0.61 & {8.12e-8}\\
                         &4 & 8.63 &	0.90 & {9.14e-8}\\
                         &5 & 8.70 &	0.91 & {7.17e-8}\\
                         &6 & 9.93	& 1.04 & {1.30e-7}\\
\end{tabular}
\end{ruledtabular}
\end{table}

To understand the excitation of different eigenstates and the overall time dynamics, Fig. 3 shows ${\omega }^{''}_{n}$ for $n=1,...,6$ as a function of $J$ for 20 nm and 80 nm cells. When $J>J_{c1}$ but smaller than the rest of critical currents, only
${\omega }^{''}_{1}<0$, whereas the rest ${\omega }^{''}_{n}>0$. As a result, only the $n=1$ eigenstate is important for the time dynamics. On the other hand, for large $J$, all ${\omega }^{''}_{n}<0$ and they are close to each other (see Eq. (15)). This behavior can be explained by noting that for large $J$, ${\omega }^{''}_{n}\approx -\alpha {J}/{{{k}_{n}}}$, so that assuming that $k_n$ is close for different $n$, ${\omega }^{''}_{n}$ is also approximately the same for different $n$. As a result, many eigenstates become important to describe the time dynamics.
\begin{figure}[htb]
\includegraphics[width=\linewidth]{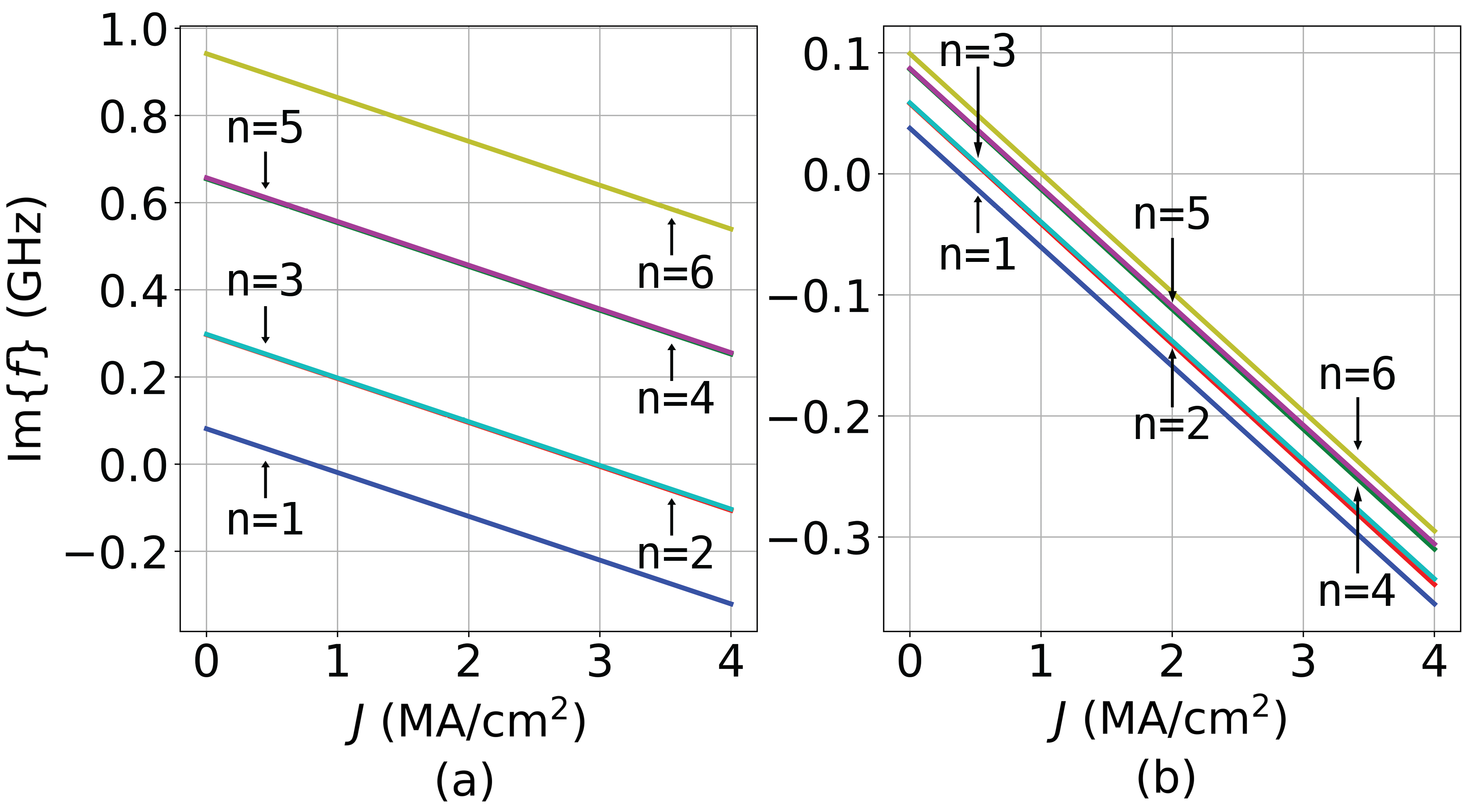}
\caption{Imaginary eigen frequencies ${\omega }^{''}_{n}$ vs. current density $J$ for the first 6 eigenstates for (a) $D= $20 nm  and (b) $D=$80 nm.}
\end{figure}

Next, Fig. 4 shows $J_{sw}$ as a function of $1/\tau$ obtained via the EVF analysis and via the complete LLGE solver. In the LLGE simulations, switching is defined as the average perpendicular magnetization crossing zero. {In EVF, the results were obtained via Eq. (25), where $\zeta$ was chosen such that curve of $J$ vs. $1/t_{sw}$ obtained for $J>J_c$  is accurately extrapolated to $J_c$  at $1/t_{sw}=0$. This choice resulted in $\zeta=1.08$ and $\zeta=1.32$ for $D=20$ nm and $D=80$ nm, respectively.} {The initial magnetization conditions for both EVF and LLGE simulations were the same equilibrium state, i.e., for EVF $\mathbf{v}(t=0)=0$ and ${{a}_{n}}(0)=<\mathbf{v}(t=0),{{A}_{0}}{{\bar{\varphi }}_{n}}>=0$.} The results obtained via the EVF and LLGE approaches are close to each other. The curve obtained via the LLGE solver is linear for the 20 nm case, but it is not linear for the 80 nm case. For the 80 nm case, the curve has different curvatures for larger and small $1/\tau$. The values of $J_c$  are typically obtained by linearly extrapolating from the $J_{sw}$ vs. $1/\tau$ curve such that ${{J}_{c}}={{J}_{sw}}({1}/{\tau }=0)$. The non-linearity of the $J_{sw}$ vs. $1/\tau$ curve, therefore, poses a significant problem in terms of the reliability and speed of calculating  $J_c$. Indeed, one needs to run simulations for a large $\tau$ to obtain more reliable results, which is slow, and it is not clear {a priori} what values of  $\tau$ are required. EVF, on the other hand, allows obtaining the results by simply obtaining a solution to a single eigenproblem problem followed by the perturbation theory analysis for the results in Table 1.
\begin{figure}
\includegraphics[width=\linewidth]{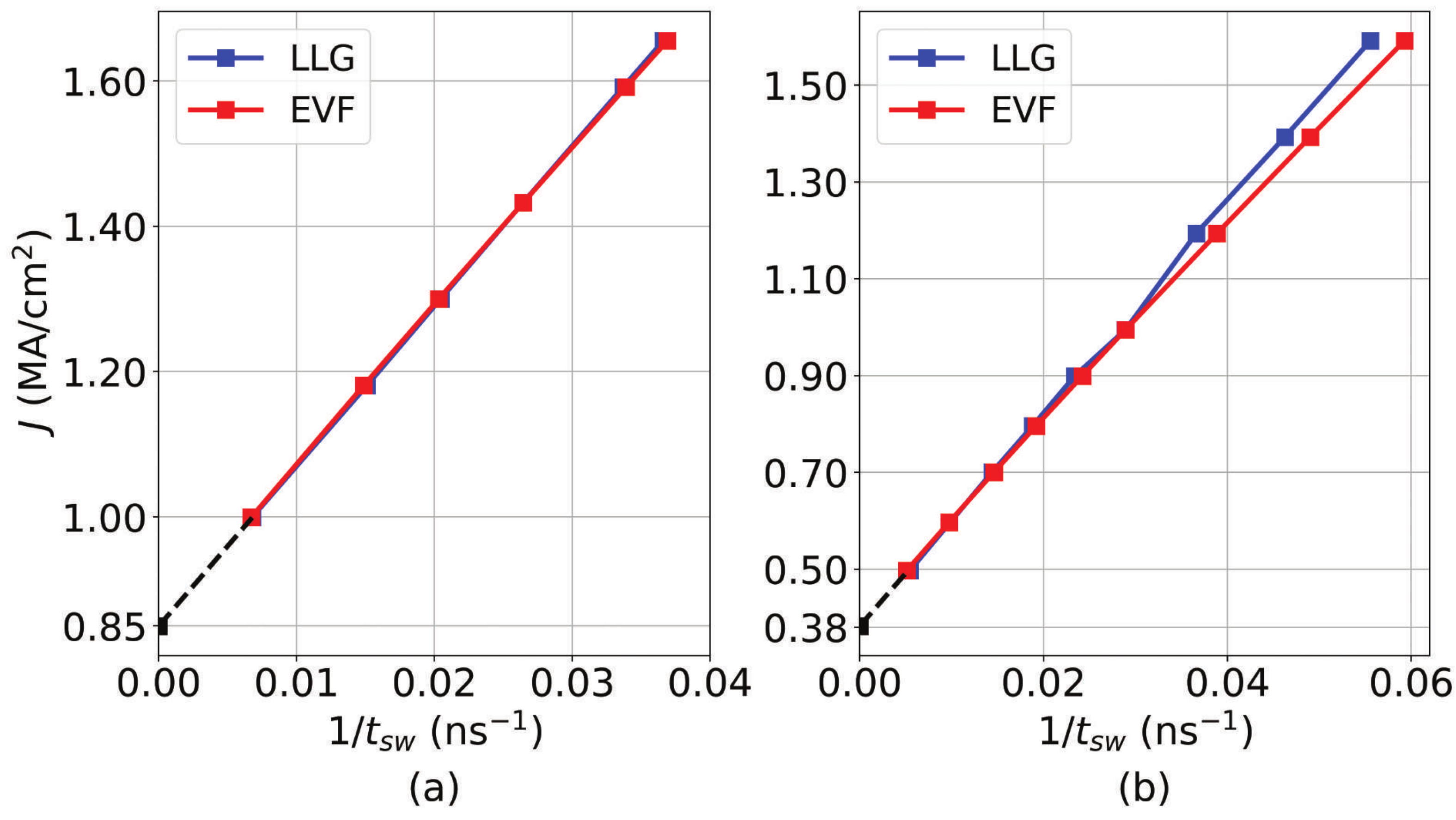}
\caption{Inverse of switching time with different current densities for (a) $D= $ 20 nm and (b) $D =$ 80 nm. Black dash lines show linear extrapolation of 5 points with smallest current of LLGE simulations and the intercept is the predicted $J_c$.}
\end{figure}

Finally, Fig. 5 demonstrates the time domain dynamics using the EVF and the LLGE solver {for the 20 nm and 80 nm cells for two values of $J$. The EVF results are shown for the overall solution $\mathbf{v}$ and for scaled $a_n$ corresponding to individual eigenstates}. The LLGE results are shown for the magnitude of the spatially averaged transverse magnetization component {${\textbf{m}}_{r}=\mathbf{m}-(\mathbf{m}\cdot {{\mathbf{m}}_{0}}){{\mathbf{m}}_{0}}$} {equivalent to $\mathbf{v}$ in EVF}. {The results are shown for the magnitude of the averages $|\langle\textbf{m}_r\rangle|$ and $|\langle\textbf{v}\rangle|$ , which represent a more global characterization, e.g., related to magnetoresistance that would be obtained if a read layer were added to the stack, as well as the average of the magnitudes $\langle|\textbf{m}_r|\rangle$ and $\langle|\textbf{v}|\rangle$, which represent more local behavior of the magnetization.} The initial magnetization conditions for both EVF and LLGE simulations were the same equilibrium state, as in Fig. 4. For EVF, we stopped the simulation when $|\langle\mathbf{v}\rangle| = 1$ , which is the physically maximal possible value. {For the 20 nm cell (Figs. 5(a, b, e, f)), the EVF results for $|\langle\mathbf{v}\rangle|$ and $\langle|\mathbf{v}|\rangle$ are close to the LLGE results $|\langle\mathbf{m}_r\rangle|$ and $\langle|\mathbf{m}_r|\rangle$ for all times until switching occurs.} The increase of {$|\mathbf{v}|$} is exponential and the increase rate is directly given by $\omega^{''}_{1}$. {This behavior is explained by the fact that for the 20 nm cell with $J=1.5J_{c1}$, $J>J_{c1}$ but $J<J_{cn}$ with $n>1$ (see Table 1), i.e., only $n=1$ eigenstate is growing with $\omega_1^{''}<0$, whereas all other eigenstates are damped with $\omega_n^{''}>0$ (Figs. 5(i, j)). For the 20 nm cell with  $J=10J_{c1}$,  $J>J_{cn}$ and $\omega_n^{''}<0$ for $n<6$, but $\omega_1^{''}$ is significantly more negative, such that the  $n=1$ eigenstate is still dominant.} For the 80 nm cell, ${{m}_{r}}$  has a similar behavior for the smaller $J$ (Figs. 5 (c, g, k)), which has the same explanation as the cases for the 20 nm cell. {The behavior of $\mathbf{m}_r$ after switching starts is more complicated because switching for the 20 nm cells is mostly by uniform rotation and for 80 nm cells it is by domain wall. For greater $J$  for 80 nm cells(Figs. 5(d)), the time dependence of $|\langle\mathbf{m}_r\rangle|$ and $|\langle\mathbf{v}\rangle|$ still appears to be mostly as an exponential increase.} On the other hand, the time dependence of {$\langle|\mathbf{m}_r|\rangle$ and $\langle|\mathbf{v}|\rangle$} (Fig. 5(h)) is not just an exponential increase but rather it is modulated with oscillations. The agreement between the eigenvalue and LLGE solver frameworks is still good for times until switching starts. {The oscillatory behavior in Fig. 5(h) is explained by the fact that multiple eigenstates become growing, and their coupling needs to be accounted for. Specifically, from Fig. 5(l) and Table 1, the $n=2$ eigenstate has a significant contribution. Because of the $n=2$ eigenstate symmetry, this contribution is not revealed in the $|\langle\mathbf{m}_r\rangle|$  and $|\langle\mathbf{v}\rangle|$  but it leads to oscillations in  $\langle|\mathbf{m}_r|\rangle$  and  $\langle|\mathbf{v}|\rangle$.} Additionally, for large amplitudes of ${{\mathbf{m}}_{r}}$, the dynamics obtained via the general LLGE solver becomes highly non-linear, e.g., the final switching may be via domain walls. Still, the EVF predicts the initial dynamics and onset of switching accurately even for such large cells.
\begin{figure}[htb]
\includegraphics[width=\linewidth]{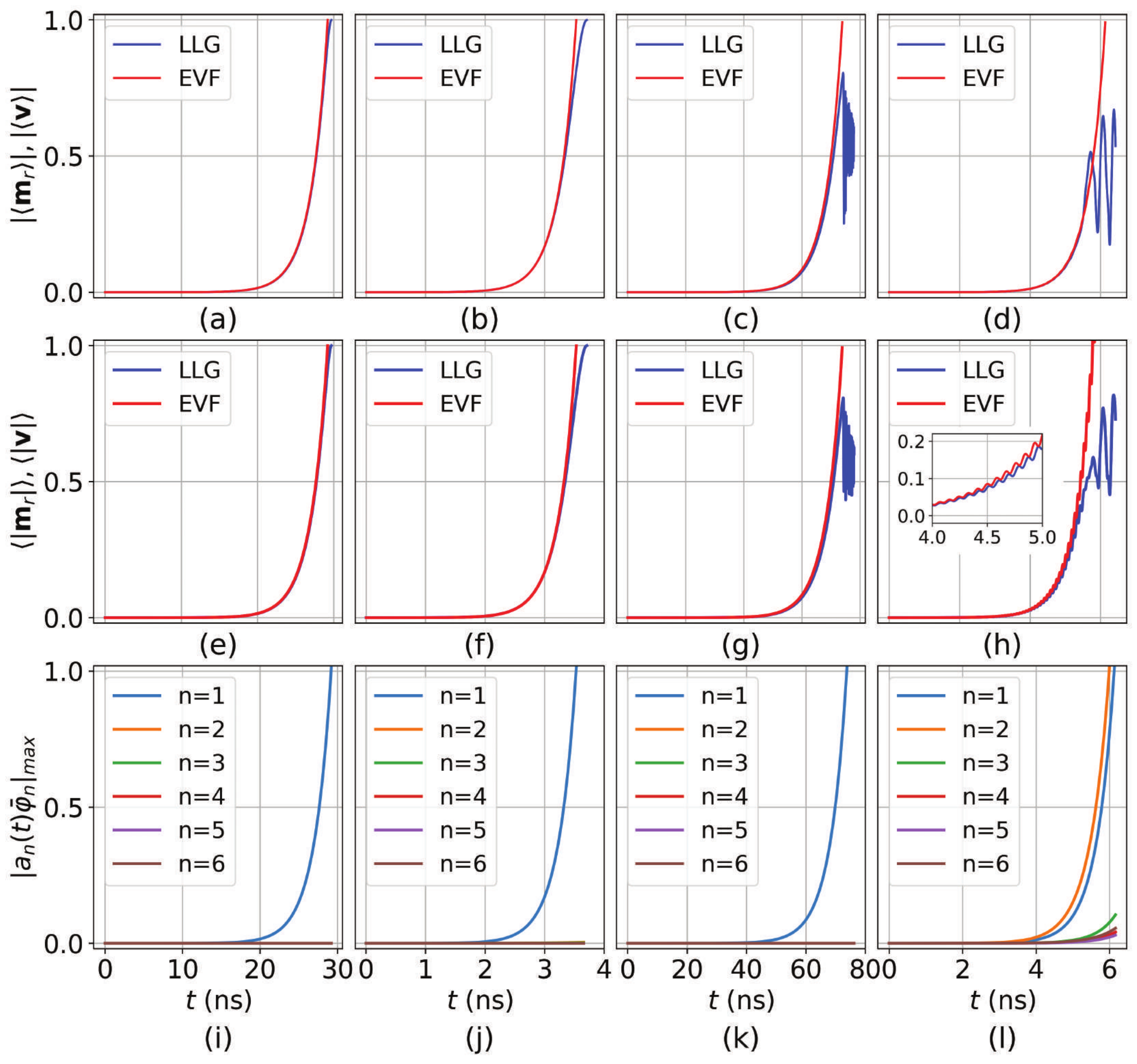}
\caption{{Time dependence of the magnetization behavior for different $D$ and different values of constant and uniform $J$. (a-d) Magnitude of the averaged $|\langle\mathbf{m}_r\rangle|$  and $|\langle\mathbf{v}\rangle|$; (e-h) Average of the magnitude of $\langle|\mathbf{m}_r|\rangle$  and  $\langle|\mathbf{v}|\rangle$; (i-l) coefficients $a_n(t)$ scaled with $|\bar{\varphi}_n|_{\max}$. The results are given for (a, e, i) $D = 20$ nm with $J=1.5J_c$; (b, f, j) $D=20$ nm  with $J=10J_c$; (c, g, k) $D=80$ nm  with $J=1.5J_c$; (d, h, l) $D=80$ nm  with $J=10J_c$. The inset in (h) presents a zoom in showing the magnetization oscillations appearing due to the excitation of multiple eigenstates.}}
\end{figure}

We note that the presented EVF is related to the micromagnetic spectral mapping technique (MSMT), which uses the Fourier transform to study spectral excitations and eigenstates in micromagnetic systems \cite{30}. Using the eigenstates provides similar information about the mode that can be extracted from MSMT. However, the eigenstates, which represent spectral information, are used to allow computing and understanding the time domain behavior, whereas in MSMT the time domain behavior is used to provide spectral information.
\section{\label{sec:5}Summary}
We presented a theoretical and numerical framework for studying the switching properties of nanomagnetic structures driven by spintronic excitations. The framework considers a linearized LLGE for the small magnetization deviations from the equilibrium state. It expands the small magnetization deviations in terms of eigenstates with corresponding complex eigenfrequencies. Depending on the current driving spin torque, the eigenfrequencies can have a positive or negative imaginary part corresponding to damped or growing time domain solutions, respectively. The system time dynamics is then driven by a small number of growing eigenstates and for small currents just a single eigenstate may be sufficient. We developed a perturbation theory that provides semi-analytical dynamic solutions by using the base eigenvalue solutions, i.e., eigenvalue problem solutions with no current or damping. The framework allows obtaining accurate predictions of the switching properties, including the critical switching current, switching time for a given current and switching current for a given time. The critical switching current is obtained as the smallest current leading to vanishing imaginary part of the eigenfrequencies. The switching time and switching current can be obtained based on the values of the imaginary part of the eigenfrequencies. {The presented EVF can also be extended to account for thermal effects.} The approach provides important insights into dynamics in such systems and allows solving several difficulties in their modeling, such extracting the switching current in MRAM and understanding reasons for switching mechanisms. The introduced framework is intended for applications in design and modeling of spintronic devices and understanding physics of their switching mechanisms.

\begin{acknowledgments}
This work was supported as part of the Quantum-Materials for Energy Efficient Neuromorphic-Computing (Q-MEEN-C), an Energy Frontier Research Center funded by the U.S. Department of Energy, Office of Science, Basic Energy Sciences under Award No. DE-SC0019273. This work used the XSEDE \cite{28}, which is supported by NSF grant number ACI-1548562, specifically, it used the Bridges and Comet systems supported by NSF Grant \# ACI-1445506.
\end{acknowledgments}


\bibliography{eigen_ref}

\end{document}